\documentclass[twoside,english]{article}
\usepackage[latin1]{inputenc}
\usepackage{pifont}
\usepackage{amssymb}

 \makeatletter

 \def\beee{\begin{equation}}
 \def\eeee{\end{equation}}

 \newcommand{\lyxaddress}[1]{
 \par {\raggedright #1
 \vspace{1.4em}
 \noindent\par}
 }

 \AtBeginDocument{

 }

 \usepackage{babel}
 \makeatother
 
\begin{document}

 \title{{\normalsize CAUSALITY IN NON-COMMUTATIVE QUANTUM FIELD THEORIES}}

 \author{Asrarul Haque\footnote{email address:  ahaque@iitk.ac.in}  and Satish D. Joglekar\footnote{email address: sdj@iitk.ac.in}}

 \maketitle

 \lyxaddress{Department of Physics, I.I.T. Kanpur, Kanpur 208016 (INDIA)}

 \begin{abstract}
We study causality in non-commutative quantum field theory with a space-space non-commutativity. We employ the S-operator approach of  Bogoliubov-Shirkov(BS). We generalize the BS criterion of causality  to the noncommutative theory. The criterion to test causality leads to a nonzero difference between T*-product and T-product as a condition of causality violation for a spacelike separation. We discuss two examples; one in a scalar theory and one in the Yukawa theory. In particular, in the context of a non-commutative Yukawa theory, with the interaction Lagrangian $\overline{\psi}(x)\star\psi(x)\star\phi(x)$, is observed to be causality violating even in case of space-space noncommutativity for which $\theta^{0i}=0$.
 \end{abstract}

 \section{Introduction}
 Nonlocal field theories, in a variety of forms, have been proposed  from time to
 time as a possible remedy against the UV divergences that arise due to the
 ill-defined product of the fields at an identical space-time point. Noncommutative
 space was  first introduced, with a similar goal, by Snyder~\cite{sny}. Later,
 noncommutative spaces were found to arise in several different contexts. Interplay
 between quantum theory and gravitation suggests a non-trivial structure of
 space-time  at short-distances and a noncommutative structure of space-time is a
 possibility. Indeed, the notion of space-time as a $ c^\infty $ manifold may not
 exist down to the distance-scales of the order of Plank length scale~\cite{dop}.
 Space-time noncommutativity naturally appears as a low energy limit of the open
 string theory on a D-brane configuration in a constant $B$-field
 background~\cite{sei}.

  We shall deal with noncommutative field theories defined on noncommutative
 manifold obeying \beee [\hat{x}^{\mu}, \hat{x}^{\nu}]= i\theta^{\mu \nu} \label{cr}
 \eeee Where $\theta^{\mu \nu}$  is a constant real antisymmetric matrix of length
 dimension two\footnote{We shall not necessarily commit ourselves to a
 value/scale for the parameter $\theta$, but leave it a free parameter to be
 determined experimentally.}. Given a local field theory on a commutative
 space-time, it  can be generalized to a noncommutative space-time. In the net effect,
 it amounts to a replacement of ordinary local product by a Moyal star product of
 the two functions~\cite{sz} \beee
  (A*B)(x): = e^{\frac{i}{2}\theta ^{\mu \nu } \partial _\mu
 ^x
 \partial _\nu ^y } A(x)B(y)\left| {_{y = x} } \right .
 \eeee

 Noncommutative quantum field theory (NCQFT) is constructed out of the action
 with the usual product of the fields replaced by the Moyal star product.
 Nonlocality of the theory stems from the Moyal product which consists of a tower
 of an arbitrarily high number of space-time derivatives. The nonlocality enters
 through only the interaction terms, because at the quadratic level both the
 commutative and noncommutative theories are identical. This is reflected from
 the fact that Moyal star product of the two fields can always be decomposed into
 the usual product of the fields and a total derivative term which would be zero if
 integrated over all space-time. Non-commutative quantum field theory \cite{sz},
 (NCQFT), is in effect, a \emph{nonlocal} quantum field theory, as is especially
 obvious from its star-product formulation. A typical non-local QFT has its
 interaction spread over a finite region at a given instant and thus this includes
 points that are separated by a space-like separation. This makes possible for a
 violation of causality in such theories. The violation of causality has been a
 subject of much discussion.
 Definitions of causality can be/have been attempted at various levels:
 \begin{itemize}
 \item The primarily meaningful definition of causality violation is linked
 with its experimental observation.
 \item Causality can also be tested in a physical situation by whether the physical cause precedes the physical effect  \cite{nsei}.
 \item However, alternate definitions are often given in terms of the \emph{micro-causality}:
 i.e. whether a given commutator (or a related object) of "local" observables vanishes outside
 the light cone~\cite{CH02,GR06,zhe}.
 \item There have been attempts to link causality with dispersion relation
 approach also for NCQFT.\cite{dis}
  \end{itemize}
For a different perspective on causality, however, see e.g. \cite{cal}\\
 In the context of a non-commutative quantum field theory, (NCQFT),
 there are no (strictly) local observables as elucidated below. Hence,
 the question of definition of micro-causality is a moot question and
 indeed various definitions of micro-causality itself have been proposed
 in the context of a NCQFT.

Consider first a  \emph{local} field theory. Suppose, we are given a local
observable, $O\left[\phi\left(x\right)\right]$ in fields (generically denoted by
$\phi$), and their \emph{finite} order time-derivatives. In a NCQFT, it will be
represented by a star product of the fields and their derivatives. For example,
$O_{1}\left[\phi\right]=\phi^{3}\left(x\right)$ will be represented by an
observable
$O_{1}^{*}\left[\phi\right]=\phi\left(x\right)\star\phi\left(x\right)\star\phi\left(x\right)$
and $O_{2}\left[\phi\right]=\phi\partial_{\mu}\phi+\partial_{\mu}\phi\phi$ will be
represented by an observable
$O_{2}^{*}\left[\phi\right]=\phi\left(x\right)\star\partial_{\mu}\phi\left(x\right)+\partial_{\mu}\phi\left(x\right)\star\phi\left(x\right)$.
For the sake of brevity and convenience, we shall call these also as {}``local'' (in
quotes) in the context of a NCQFT .

We first enumerate below the definitions of micro-causality
suggested :

\begin{enumerate}
\item For two {}``local'' observables $O{}_{1}^{*}\left(x\right)$ and
$O_{2}^{*}\left(y\right)$, the theory violates micro-causality \cite{CH02} if
$\left[O_{1}^{*}\left(x\right),O_{2}^{*}\left(y\right)\right]\neq0,$ for
$\left(x-y\right)^{2}<0$. Here,
$\left[O{}_{1}^{*}\left(x\right),O_{2}^{*}\left(y\right)\right]$ stands for the
\emph{commutator} . We shall use $x\sim y$ to imply that $x$ and $y$ are space-like separated.
\item For two {}``local'' observables $O_{1}^{*}\left(x\right)$ and $O_{2}^{*}\left(y\right)$,
the theory violates micro-causality if
$\left[O_{1}^{*}\left(x\right),O_{2}^{*}\left(y\right)\right]_{*}\neq0,$ for
$\left(x-y\right)^{2}<0$. Here,
$\left[O_{1}^{*}\left(x\right),O_{2}^{*}\left(y\right)\right]_{*}$ stands for the
\emph{star-commutator} \cite{GR06} defined by: \[
\left[O_{1}^{*}\left(x\right),O_{2}^{*}\left(y\right)\right]_{*}=O{}_{1}^{*}\left(x\right)\star
O{}_{2}^{*}\left(y\right)-O{}_{2}^{*}\left(y\right)\star O{}_{1}^{*}\left(x\right)\] with
\[ O_{1}^{*}\left(x\right)\star O{}_{2}^{*}\left(y\right)\equiv exp\left\{
\frac{i}{2}\theta^{\mu\nu}\partial_{\mu}^{x}\partial_{\nu}^{y}\right\}
O_{1}^{*}\left(x\right)O{}_{2}^{*}\left(y\right)\] for an \emph{arbitrary} pair of points $x,y$.
\end{enumerate}
In this work, we shall attempt to look at the problem of causality from another perspective. This is based on the approach by Bogoliubov and Shirkov [BS] \cite{bs} . They have formulated a general $S-$ operator approach (however, for a commutative space-time) that does not require one to commit to a specific field theory setting and is based upon a primary definition of causality: a physical disturbance cannot propagate out of its forward light-cone. This approach thus has a direct physical basis and has been found useful in non-local quantum field theories \cite{jj04}. We generalize this approach, as far as it is possible, to a space-space NCQFT and develop a criterion based on this approach to test causality.  In section 2, we first summarize works related to the causality. Generalization of the BS-approach requires that we introduce a space-time dependent coupling in the intermediate stages. In section 3, we first carry out this generalization. Another issue that differs from the BS approach to commutative field theory is that the use of space-like intervals needs a reformulation. In this section, we go over the argument for BS criterion for a NCQFT to see where it needs a revision. In section 4, we arrive at a criterion to test causality. In the section 5, we work out two examples, one in the scalar NCQFT  and another in the Yukawa NCQFT for causality violation (CV). We show that in either cases, even for space-space non-commutativity, there is CV. In section 6, we shall connect the causality criterion to the compatibility of measurement process for two "local" observables.

\section{Summary of Earlier Works}
Should micro-causality be valid for a theory, we expect that any pair of observables, $O_1^ * (x)$ and $O_2^*(y)$, should commute for a spacelike separation \beee [ {O_1^ * (x),O_2^*(y)} ] = 0 ~~~~ {\rm{ for~~
x}} \sim y \label{eq:mc}\eeee\hspace{-.05in}
where $O_{1,2}^*(x)$ stand for  "local" observables in a noncommutative
 theory in which fields or derivative of fields are combined via the Moyal star product (such as~~$\phi(x)\star\phi(x)\star\phi(x))$. In particular, we expect that (\ref{eq:mc}) should hold as an operator equation, i.e. \emph{every} matrix element of $[ {O_1^ * (x),O_2^*(y)} ]$ should vanish in such a case.\\

Original works are based on the above definition of
 micro-causality. Chaichian, et al,~\cite{CH02} observed that the
 matrix element$$
 \left\langle {\rm{0}} \right|\left[ {{\rm{:\phi (x)\star\phi(x)}:,\rm{:\phi (y)\star\phi(y):}}}
 \right]_{{\rm{x}}_{\rm{o}} {\rm{ = y}}_{\rm{o}} } \left| {{\rm{p,p'}}} \right\rangle$$
  is nonzero for $\theta ^{0i}  \ne 0$ and thus
 violates micro-causality in the space-time NCQFT. They also generalized the idea to the case of  field theories with light-like noncommutativity: $\theta^{\mu\nu}\theta_{\mu\nu}=0$.\\

 On the other hand, Greenberg~\cite{GR06}
 has calculated the following matrix element of the following commutator  $$ \left\langle 0 \right|\left[ :{\varphi (x) \star \varphi (x):,\partial _0 \left(:\varphi (y) \star \varphi (y):\right)} \right]_{x_o  = y_o } \left| {p,p'} \right\rangle
$$ and drew attention to the fact that it fails to obey micro-causality even for the case $\theta ^{0i} = 0$ i.e. for the space-space noncommutativity.

Greenberg then introduced the Moyal (star) commutator which reads \beee
 [A,B]^* = A\star B - B\star A
\eeee \hspace{-.05in}
 and analyzed the quantity $
 \left\langle {0\left| {\left[ {{\rm{:\phi (x)\star\phi (x):,:\phi (y)\star\phi (y):}}}
 \right]^*_{{\rm{x}}_{\rm{0}} {\rm{ = y}}_{\rm{0}} } } \right|p,p'} \right\rangle
  $ which  violates micro-causality even in case of space-space
noncommutativity in which $\theta ^{0i} = 0$. He noted that the star commutator, unlike the ordinary commutator, is sensitive to the separation of $x$ and $y$ through Moyal phases and suggested that it is the star commutator, and not the commutator,  that is relevant for microcausality.

Zheng~\cite{zhe} has studied the star commutator further. He has calculated the vacuum expectation value of equal time star commutator\\ $ \left\langle {0\left| {\left[ {:\phi (x)\star\phi (x):,:\phi (y)\star\phi (y):}
 \right]^*_{x_0  = y_0 } } \right|0} \right\rangle$ which vanishes for $\theta ^{0i} = 0$.
Zheng also studied the vacuum and the non-vacuum matrix elements of the quantity
$ {\left[ {:\bar \psi _\alpha  (x)\star\psi _\beta  (x):,:\bar \psi _\sigma  (y)\star\psi _\tau  (y):}\right]^* _{x_0 = y_0 } } $ and showed that it does not vanish  for spacelike separation, no matter whether $\theta ^{0i} =
 0$ or $\theta ^{0i} \ne 0$.

\section{Development of Criterion for Causality\label{BS}}

We shall first develop a criterion for causality violation (CV) along the lines of
\cite{bs} Bogoliubov-Shirkov (BS), appropriately generalized to a
non-commutative (NC) space-time. This will also enable us to construct a
quantity that will enable us to decide under what conditions are two observables
compatible (as further discussed in section 6).
We shall restrict ourselves to a non-commutative space-time with
$\theta^{12}=-\theta^{21}\equiv \theta\neq0$ and $\theta^{\mu\nu}=0$ otherwise.
In this frame of reference, time $t$ is well-defined and this makes a
generalization of the BS criterion easier for such NC quantum field theories.

The BS discussion begins with an S-operator. For the formulation of the BS
criterion, we need a variable coupling $g(x)$ that can be varied over the
space-time; and the S-operator, $S\left[g\right]$, for such a coupling\footnote{The
idea of a variable coupling, at least over time, is not new: it is employed in the
LSZ formulation.}. Before we proceed with the generalization for the case of
NCQFT, we shall first generalize the interaction term to include a variable
$g\left(x\right)$.

\subsection{Interaction Term\label{subsec:IT}}

Let the interaction of a local commutative field theory be

\[
S_{I}=g\int d^{4}x\mathcal{L}_{I}\left[\phi\left(x\right)\right]\] In the space-time
dependent coupling formalism, it would be replaced by \[ S'{}_{I}=\int
d^{4}xg\left(x\right)\mathcal{L}_{I}\left[\phi\left(x\right)\right]\] In a non-commutative
space-time, this would be replaced by,\[ S_{I}\rightarrow\int
d^{4}xg\left(x\right)\star\mathcal{L}_{I}^{*}\left[\phi\left(x\right)\right]\] Here,
$\mathcal{L}_{I}^{*}\left[\phi\left(x\right)\right]$ is a short-hand notation for the
interaction $\mathcal{L}_{I}\left[\phi\left(x\right)\right]$ converted to a
non-commutative space star-product. For example, $S'{}_{I}=\int
d^{4}xg\left(x\right)\phi^{4}\left(x\right)$ would be replaced by \[ \int
d^{4}xg\left(x\right)\phi^{4}\left(x\right)\rightarrow\int
d^{4}xg\left(x\right)\star\phi\left(x\right)\star\phi\left(x\right)\star\phi\left(x\right)\star\phi\left(x\right)\]
It is easy to verify however that,\begin{equation} \int d^{4}x\;
g\left(x\right)\star\phi\left(x\right)\star\phi\left(x\right)\star\phi\left(x\right)\star\phi\left(x\right)\equiv\int
d^{4}x\;
g\left(x\right)\phi\left(x\right)\star\phi\left(x\right)\star\phi\left(x\right)\star\phi\left(x\right)\label{eq:starless}\end{equation}
To see this, consider the expression on the left hand side of (\ref{eq:starless}),
written in momentum space.\begin{eqnarray*}
 &  & \int d^{4}xg\left(x\right)\star\phi\left(x\right)\star\phi\left(x\right)\star\phi\left(x\right)\star\phi\left(x\right)\\
 & = & \int d^{4}kd^{4}k_{1}d^{4}k_{2}d^{4}k_{3}d^{4}k_{4}exp\left\{ \frac{-i\theta^{\mu\nu}\left[k_{\mu}\left(k_{1}+k_{2}+k_{3}+k_{4}\right)_{\nu}+O.T.\right]}{2}\right\} \\
 & \times & g\left(k\right)\phi\left(k_{1}\right)\phi\left(k_{2}\right)\phi\left(k_{3}\right)\phi\left(k_{4}\right)\delta^{4}\left(k+k{}_{1}+k_{2}+k_{3}+k_{4}\right)\\
 & = & \int d^{4}kd^{4}k_{1}d^{4}k_{2}d^{4}k_{3}d^{4}k_{4}g\left(k\right)exp\left\{ \frac{-i\theta^{\mu\nu}\left[O.T.\right]}{2}\right\} \\
 & \times & \phi\left(k_{1}\right)\phi\left(k_{2}\right)\phi\left(k_{3}\right)\phi\left(k_{4}\right)\delta^{4}\left(k+k_{1}+k_{2}+k_{3}+k_{4}\right)\\
 & = & \int d^{4}xg\left(x\right)\phi\left(x\right)\star\phi\left(x\right)\star\phi\left(x\right)\star\phi\left(x\right)\end{eqnarray*}
In the second line, $O.T.$ stands for other terms in the exponent not containing
$k$ and in third line, we have used $\theta^{\mu\nu}k_{\mu}k_{\nu}\equiv0$. Thus,
for example, $\frac{\delta S_{I}}{\delta
g\left(x\right)}=\phi\left(x\right)\star\phi\left(x\right)\star\phi\left(x\right)\star\phi\left(x\right)$.
We note that the S-matrix in the lowest nontrivial order is $gS_{1}$and is entirely
generated by the tree-order matrix elements of $S_{I}$ .Unitarity of the S-matrix
to this order implies,\[ (1+gS_{1})^{\dagger}(1+gS_{1})=1+O\left(g^{2}\right)\]
which leads to \[ S_{1}^{\dagger}=-S_{1}=-iS_{I}\]

\subsection{Space-Like Intervals\label{subsec:LT}}
In the discussion of the BS criterion of causality for a commutative space-time,
use is often made of space-like intervals: $(x-x')^2<0$. On the noncommutative
spaces in question, the $x-y$ coordinates do not commute. As such, we need to
consider $(x-x')^2$ as an operator. We can still consider two space-time points
which are specified by \emph{definite} values both for $x_0,x_3$ and
$x'_0,x'_3$. However, $X\equiv x-x'$ and $Y \equiv y-y'$ are both operators. It is
not difficult to see that
\begin{eqnarray}
  X^2+Y^2 &=&(X+iY)(X-iY) +i\left[X,Y\right] \nonumber \\
  &=& (X-iY)(X+iY) -i\left[X,Y\right]  \nonumber \\
   <X^2+Y^2 >&\geq&\left| \left[X,Y\right]\right| \nonumber \\
   &=&2\theta
\end{eqnarray}
in view of the positive semi-definiteness of the operators $(X+iY)(X-iY)$ and
$(X-iY)(X+iY) $. Thus, we shall regard an interval $(x-x')$ space-like if
$(x_0-x'_0)^2-(x_3-x'_3)^2<2\theta$. \\
\begin{itemize}
 \item For a space-like separation  with $(x_0-x'_0)^2-(x_3-x'_3)^2<0$, it is possible to change the order of
time coordinates $x_0$ and $x'_0$ by a Lorentz transformation confined to
the $z-t$ plane alone. Such Lorentz transformations preserve the nature of
non-commutativity (i.e. space-space). We shall call such a space-like
separation a 'restricted' one and shall denote it by $x\asymp y$.
\item Two distinct events $x,x'$ with $x_0=x'_0$ and $x_3=x'_3$ can always be enclosed in  \emph{some} disjoint neighborhoods in this plane, compatible with $\Delta x_1\Delta x_2 =\frac{1}{2}\theta$. Hence, a theory cannot be causal if it necessarily allows instantaneous propagation of a signal from one to another.
\end{itemize}
We shall see, in section 4, that the criterion of causality demands that the commutator of the interaction Lagrangian vanish over just the two sets of points we have discussed.
\subsection{Generalization of BS criterion  to a NCQFT\label{subsec:gen}}

Let $\left\{ \left|\alpha,in\right\rangle \right\} $ denote a complete
set of scattering in-states. We shall consider a particular matrix
element

\[
S_{\beta\alpha}=\left\langle \beta,in\right|S[g]\left|\alpha,in\right\rangle \]
For a constant $g$, $S_{\beta\alpha}$ has the perturbative expansion:\[
S_{\beta\alpha}=\delta_{\beta\alpha}+gS_{\beta\alpha}^{(1)}+\frac{g^{2}}{2!}S_{\beta\alpha}^{(2)}+........\]
and for a variable $g\left(x\right)$, it has an expansion:\begin{equation}
S_{\beta\alpha}\left[g\right]=\delta_{\beta\alpha}+\int d^{4}xg\left(x\right)S_{\beta\alpha}^{(1)}\left(x\right)+\frac{1}{2!}\int d^{4}xd^{4}yg\left(x\right)g\left(y\right)S_{\beta\alpha}^{(2)}\left(x,y\right)+....\label{eq:Sexp}\end{equation}
We want to generalize this to the non-commutative quantum field theories.
We expect the second term on the right hand side to be replaced by%
\footnote{We recall that in a QFT, $S^{(1)}$is a field operator.%
} (see also the subsection \ref{subsec:IT})
\begin{eqnarray*}
\int d^{4}xg\left(x\right)S_{\beta\alpha}^{(1)}\left(x\right)& \rightarrow & Tr \left[ g\left(\widehat{x}\right)S_{\beta\alpha}^{(1)}\left(\hat{x}\right)\right]=\int d^{4}xg\left(x\right)\star S_{\beta\alpha}^{(1)*}\left(x\right)\\
& \equiv & \int d^{4}xg\left(x\right)S_{\beta\alpha}^{(1)*}\left(x\right)
\end{eqnarray*}
If we had a constant coupling, we would have replaced\[
\int d^{4}xgS_{\beta\alpha}^{(1)}\left(x\right)\rightarrow\int d^{4}xgS_{\beta\alpha}^{(1)*}\left(x\right)\]
i.e. the replacements in the two case are identical: $S_{\beta\alpha}^{(1)}\left(x\right)\rightarrow S_{\beta\alpha}^{(1)*}\left(x\right)$.
In a similar manner, the third term on the right hand side\begin{eqnarray*}
 &  & \frac{1}{2!}\int d^{4}xd^{4}yg\left(x\right)g\left(y\right)S_{\beta\alpha}^{(2)}\left(x,y\right)\\
 & \rightarrow & \frac{1}{2!}\int d^{4}xd^{4}yg\left(x\right)\star S_{\beta\alpha}^{(2)*}\left(x,y\right)\star g\left(y\right)\\
 & \equiv & \frac{1}{2!}\int d^{4}x\left\{ \int d^{4}yg\left(x\right)*\left.S^{(2)*}_{\beta\alpha}\left(x,y\right)exp\left\{ \frac{i\theta^{\mu\nu}\partial_{\mu}^{y}\partial_{\nu}^{y_{1}}}{2}\right\} g\left(y_{1}\right)\right|_{y=y_{1}}\right\} \end{eqnarray*}
We can now carry out the integration over $y$ for a fixed $x$ and
find that the non-commutative phase cancels out. In a similar manner
one can deal with the $x$-integration and find that\begin{eqnarray*}
\frac{1}{2!}\int d^{4}xd^{4}yg\left(x\right)g\left(y\right)S_{\beta\alpha}^{(2)}\left(x,y\right)\\
\rightarrow\frac{1}{2!}\int d^{4}xd^{4}yg\left(x\right)\star S^{(2)*}_{\beta\alpha}\left(x,y\right)\star g\left(y\right)\\
\equiv\frac{1}{2!}\int d^{4}xd^{4}yg\left(x\right)g\left(y\right)S^{(2)*}_{\beta\alpha}\left(x,y\right).\end{eqnarray*}
This can be generalized to the remaining terms in (\ref{eq:Sexp}).

Thus, in the non-commutative theory also, we have an expansion of
the same form as the commutative case:\begin{equation}
S_{\beta\alpha}\left[g\right]=\delta_{\beta\alpha}+\int d^{4}xg\left(x\right)S_{\beta\alpha}^{(1)}\left(x\right)+\frac{1}{2!}\int d^{4}xd^{4}yg\left(x\right)g\left(y\right)S_{\beta\alpha}^{(2)}\left(x,y\right)+....\label{eq:Sexp2}\end{equation}
where, we have now dropped the star on $S_{\beta\alpha}^{(n)}$as
we shall employ (\ref{eq:Sexp2}) only for the NCQFT%
\footnote{The purpose of the star on $S_{\beta\alpha}^{(n)}$was to remind us
that the $S-$operator is different for the local and the NCQFT. %
}. We shall employ henceforth.

We note in passing that it is not \emph{necessary} to employ the S-operator
($U\left(-\infty,\infty\right)$) in this formulation. This observation
becomes relevant especially for a theory for which some of the S-matrix
elements may not exist because of infrared divergences. The formulation
can alternately be given also in terms of the unitary time-evolution
operator $U\left[-T,T';g\right].$

Let us now recall that we are considering a theory on a space with
$\theta^{0i}=0$ and that the time-coordinate is well-defined and
we can order the space-time points by their time-coordinate.

The derivation of the BS condition of causality proceeds much the
same way as for the commutative space-time.

We define the coupling constant functions:\begin{eqnarray*}
g\left(x\right) & = & g_{2}\left(x\right)\qquad T'>x_{0}>0\\
 & = & g_{1}\left(x\right)\qquad0> x_{0}>-T;\\
G_{2}\left(x\right) & = & g_{2}\left(x\right)\qquad T'>x_{0}>0\\
 & = & 0,\qquad \mbox{\,otherwise};\\
G_{1}\left(x\right) & = & 0\qquad T'>x_{0}>0\\
 & = & g_{1}\left(x\right)\qquad0> x_{0}>-T.\end{eqnarray*}
Now, causality demands that the evolution for $0>x_{0}>-T$ is unaffected by the
value of the coupling for $T'>x_{0}>0$. This fact is not contradicted by the $x-y$
non-commutativity. We recall that the matrix elements of $U$ depend only on the
\emph{coupling constant function} $g\left(x\right)$ and not on
$g\left(\hat{x}\right)$. Thus, should causality hold,
\[
U\left(-T,0;g\right)=U\left(-T,0;g_{1}\right)=U\left(-T,0;G_{1}\right).\]
Also, $U\left(0,T';G_{1}\right)\equiv\mathcal{I}$. Also, the \emph{evolution}
operator for $t>0$ depends only on the coupling for $t>0$. Hence,\[
U\left(0,T';g\right)=U\left(0,T';G_{2}\right)\]
Thus,\begin{eqnarray*}
U\left(-T,T';g\right) & = & U\left(0,T';g\right)U\left(-T,0;g\right)\\
 & = & U\left(0,T';G_{2}\right)U\left(-T,0;G_{1}\right)\\
 & = & U\left(-T,T;G_{2}\right)U\left(-T,T;G_{1}\right)\end{eqnarray*}
In a similar manner, for \begin{eqnarray*}
g'\left(x\right) & = & g'_{2}\left(x\right)\qquad T'>x_{0}> 0\\
 & = & g_{1}\left(x\right)\qquad0> x_{0}>-T;\end{eqnarray*}
we have \begin{eqnarray*} U\left(-T,T';g'\right) & = &
U\left(-T,T';G'_{2}\right)U\left(-T,T';G_{1}\right)\end{eqnarray*} where we have
defined, in an analogous manner,
\begin{eqnarray*}
G'_{2}\left(x\right) & = & g'_{2}\left(x\right)\qquad T'>x_{0}> 0\\
 & = & 0,\qquad \mbox{otherwise};\end{eqnarray*}
Then,\begin{eqnarray} U\left(-T,T';g'\right)U^{\dagger}\left(-T,T';g\right) & = &
U\left(-T,T';G'_{2}\right)U^{\dagger}\left(-T,T';G_{2}\right)\label{axz}\end{eqnarray}
and is independent of values of  $g_{1}(x)$ for $-T< x_0< 0$ . This is the BS
condition of causality. We may write the above equation in the form \beee U( -
T,T';g(y) + \delta g(y))U^\dag ( - T,T';g(y)) = 1 + \delta U( - T,T';g(y))U^\dag  ( -
T,T';g(y)) \eeee where $\delta g(y) \ne 0$  for some $T' > y_0>0$. This expression needs not
depend upon the behavior of $g(x)$ for $-T< x_0<0$. So, we  have \beee
\frac{\delta }{{\delta g(x)}}\left( {\frac{{\delta U(g)}}{{\delta g(y)}}U^\dag (g)} \right) =
0~~ for~~ x < y, \label{eq:cr} \eeee ($x<y$ stands for $x_0<y_0$). This is the expression of causality in terms of
the unitary time-evolution
operator.\\
In the case of commutative QFT, the above condition also holds for $x\sim y$;
since in such a case, it is possible to make a Lorentz transformation to a frame in which $x_0<y_0$ holds. In
the present case, there is a restriction on the possible Lorentz transformation
that preserves the nature of non-commutativity. From the discussion of
subsection \ref{subsec:LT}, it follows that eq.(\ref{eq:cr}) holds also for a
'restricted' space-like separation $x\asymp x'$, i.e. with
$(x_0-x'_0)^2-(x_3-x'_3)^2<0$.\\
  Further, for two distinct points $x,y$ with $x_0=y_0=0$ and $x_3=y_3$, we note that the quantity $U( -
T,T';g(y) + \delta g(y))U^\dag ( - T,T';g(y))$, for $\delta g\neq 0$ only at $y$,  is not dependent on the value of $g$ at such an $x$ . This follows from our observation in section \ref{subsec:LT} that such points cannot be connected by a signal if causality is always to be ensured. This leads to the validity of (\ref{eq:cr}) also for such a pair of points.\\
We can express the matrix $ {\hat U}$  in the form of functionals in powers of
g(x):
\begin{eqnarray}
 U[g] &= & 1 + \sum\limits_{n \ge 1} {\frac{1}{{n!}}} \int {U_n (x_1 ,....,x_n )g(x_1 )....g(x_n )dx_1 ....dx_n } ,\nonumber \\
 &=& 1 + \int {U_1 (x_1 )g(x_1 )dx_1  + } \int {U_2 (x_1 ,x_2 )g(x_1 )g(x_2 )dx_1 dx_2  + } .....\label{eq:exp}
 \end{eqnarray}
Where $U_n (x_1 ,....,x_n )$ is a  symmetric operator with respect
to all arguments, and depends upon the field operators and on their
partial derivatives at the points $x_1 ,....,x_n$.

Unitarity of ${\hat U}$ matrix i.e. $ U^\dag[g]U[g] = 1 $ leads to
the condition, for each n, given by:
\begin{eqnarray}
U_{n}(x_{1},...,x_{n})+U_{n}^{\dagger }(x_{1},...,x_{n})\nonumber\\
 +\sum _{1\leq k\leq
{n-1}}P\left(\frac{x_{1},...,x_{k}}{x_{k+1},...,x_{n}}\right)U_{k}(x_{1},...,x_{k})U_{n-k}^{\dagger
}(x_{k+1},...,x_{n}) =  0 .
\end{eqnarray}
The symbol
 $P\left(\frac{x_{1},...,x_{k}}{x_{k+1},...,x_{n}}\right)$ stands for the sum over the distinct ways of partitioning
($\frac{n!}{k!(n-k)!} $ in number)
$\{x_{1},x_{2},x_{3},........x_{n}\}$ into two sets of $k$ and
$(n-k)$ (such as $\{x_{1},x_{2},x_{3},........x_{k}\}$
$\{x_{k+1},........x_{n}\}$).

Using(\ref{axz}), condition of causality can be expressed as:
\begin{eqnarray}
& &C_{n}(y,x_{1},...,x_{n})=iU_{n+1}(y,x_{1},...,x_{n})  \nonumber \\
& + &i\sum _{0\leq k\leq
n-1}P\left(\frac{x_{1},...,x_{k}}{x_{k+1},...,x_{n}}\right)U_{k+1}(y,x_{1},...,x_{k})U_{n-k}^{\dagger
}(x_{k+1},...,x_{n}) \nonumber  \\& =& 0\label{eq:crit}
\end{eqnarray}
Now, causality condition for n=1,2 reads as
\begin{equation}
 C_{1}(x,y)\equiv
iU_{2}(x,y)+iU_{1}(x)U_{1}^{\dagger
}(y)=0\label{causal1}\end{equation}
\begin{equation}
C_{2}(x,y,z)\equiv iU_{3}(x,y,z)+iU_{1}(x)U_{2}^{\dagger
}(y,z)+iU_{2}(x,y)U_{1}^{\dagger }(z)+iU_{2}(x,z)U_{1}^{\dagger
}(y)=0\label{causal2}\end{equation} and unitary condition gives
\begin{equation}
U_{1}(x)+U_{1}^{\dagger }(x)=0\label{unitary1}\end{equation}
 \begin{equation}
U_{2}(x,y)+U_{2}^{\dagger }(x,y)+U_{1}(x)U_{1}^{\dagger
}(y)+U_{1}(y)U_{1}^{\dagger }(x)=0\label{unitary2}\end{equation}

'S' matrix can always be recovered from the unitary time-evolution
operator in the large(infinite) time limit. So, we can have
causality and unitarity condition for n=1,2 like above as follows:

Causality condition:
\begin{equation}
 iS_{2}(x,y)+iS_{1}(x)S_{1}^{\dagger
}(y)=0\label{causal3}\end{equation}
\begin{equation}
 iS_{3}(x,y,z)+iS_{1}(x)S_{2}^{\dagger
}(y,z)+iS_{2}(x,y)S_{1}^{\dagger }(z)+iS_{2}(x,z)S_{1}^{\dagger
}(y)=0\label{causal4}\end{equation}
 Unitarity condition:
\begin{equation}
S_{1}(x)+S_{1}^{\dagger }(x)=0\label{unitary3}\end{equation}
 \begin{equation}
S_{2}(x,y)+S_{2}^{\dagger }(x,y)+S_{1}(x)S_{1}^{\dagger
}(y)+S_{1}(y)S_{1}^{\dagger }(x)=0\label{unitary4}\end{equation}
In particular, if the theory has T-invariance, the S-operator for the time-reversed theory is $S^\dagger$: $TST^{-1}=S^\dagger$. We now apply this to an analogue of (\ref{eq:exp}) for the $S$-operator and invoke\footnote{If the original theory with a constant coupling has a time-reversal invariance, the intermediate action $S[g(x)]$, with a space-time dependent coupling $g(x)$, can be made time-reversal invariant, if we associate the following transformation for $g(x)$.} $Tg(t,\textbf{x})T^{-1}=g(-t,\textbf{x})$. Then, we obtain,   \begin{equation}
S^\dagger_2(x,y)=\mathcal{I}_TS_2(-x_0,\textbf{x};-y_0,\textbf{y})\label{eq:s2dag}\end{equation}
Where, $\mathcal{I}_T$ stands for the operation of changing the sign of time labels at the end of a calculation of a matrix element. Then, (\ref{unitary4}) implies,
\begin{equation}
S_{2}(x,y)+\mathcal{I}_TS_2(-x_0,\textbf{x};-y_0,\textbf{y})+S_{1}(x)S_{1}^{\dagger
}(y)+S_{1}(y)S_{1}^{\dagger }(x)=0\label{unitary4'}\end{equation}
We shall soon demonstrate that causality implies,
\begin{equation}
S_{2}(0,\textbf{x};0,\textbf{y})=\left.\frac{1}{2}\left[S_{1}(x)S_{1}(y)+S_{1}(y)S_{1}(x)\right]\right|_{x_0=y_0=0}\label{unitary4''}\end{equation}
which is compatible with (\ref{unitary4'}) (Note: $S^\dagger_1=-S_1$).

\section{The B-S Causality Criterion}

As shown in the section \ref{BS}, the causality condition can be expressed along similar lines for a non-commutative quantum field theory as for the commutative
one. One of these is :\begin{equation} H_{1}\left(x,y\right)\equiv
iS_{2}\left(x,y\right)+iS\left(x\right)S_{1}^{\dagger}\left(y\right)=iS_{2}\left(x,y\right)-iS_{1}\left(x\right)S_{1}\left(y\right)=0,\qquad
x>\asymp y\label{eq:cc}\end{equation} \\
This implies,
 \begin{equation}
S_{2}(x,y)= S_{1}(x)S_{1}(y)=-O^*(x)O^*(y)~~~for~ x >\asymp y
 \label{unitary5}\end{equation}
since, $S_1(x)=iO^*(x)$.
If we interchange $x,y$ and use the symmetry of $S_2(x,y)$, we have
\begin{equation}
S_{2}(x,y)= S_{1}(y)S_{1}(x)=-O^*(y)O^*(x)~~~for~ y >\asymp x
 \label{unitary6}\end{equation}
We now consider two points $x,y$ such that $x\asymp y$. Then, for such a case, (\ref{unitary5}) and (\ref{unitary6}) lead to,
\begin{equation}
\left[ S_1(x),S_1(y)\right]=0\qquad\qquad x\asymp y\label{eq:com1}
\end{equation}
We shall now look at causality in a general case, i.e. we allow $x,y$ to be arbitrary (We have left out the case of $x_0=y_0$). From the remarks following (\ref{eq:cr}), we  note that (\ref{unitary5}) and (\ref{unitary6}) are  valid also when  $x_0=y_0$ for $\textbf{x} \neq \textbf{y}$ (the case with $x_3\neq y_3$ is already covered). Employing the symmetry of $S_2(x,y)$, we have,
\begin{equation} S_2(x,y)=\frac{1}{2}\left[S_1(x)S_1(y)+S_1(y)S_1(x)\right]\;\;\;x_0=y_0\;\;\textbf{x}\neq \textbf{y}\label{unitary5'}
\end{equation}
Combining (\ref{unitary5}),(\ref{unitary6}) and (\ref{unitary5'}), we have a consequence of causality
condition\footnote{We have adopted a symmetric definition for
$\theta(x_0-y_0)$:\quad$\theta(x_0-y_0)+\theta(y_0-x_0)=1\Rightarrow\theta(0)=1/2$.}
\begin{equation}
S_{2}(x,y)= i^2T[O^*(x)O^*(y)] \qquad\textbf{x}\neq \textbf{y}\label{unitary7} \end{equation}
In addition, we also have,
\begin{equation}
\left[ S_1(x_0,\textbf{x}),S_1(y_0,\textbf{y})\right]=0,\qquad x_0=y_0.\label{eq:com2}\end{equation}
The above equation is valid at the set of points characterized by $x_0=y_0,x_3=y_3$ not included in the domain of validity of (\ref{eq:com1}), viz. $x\asymp y$. We observe that the criterion of causality requires that the interaction Lagrangian $\mathcal{L}_I(x)$ commute with itself, $\mathcal{L}_I(y)$, whenever $(x-y)^2$ fulfills either of the conditions mentioned in the subsection 3.2. We also note that this consequence has \emph{followed} from the primary meaning of causality employed in subsection \ref{subsec:gen} (together with other principles). \\

Now, the equation  (\ref{unitary7}), demanded by causality, may not always be
obeyed. First we recall that when interaction term $S_{1}$ contains time
derivatives, it is well known that time ordered product in (\ref{unitary7}) is not
covariant. In QFT we often introduce another time ordered product, the $T^*-$
product which is covariant (in a commutative case). It is known that in the path integral formulation, we
naturally generate the Green's function of $T^*-$ ordered product of field
operators.  Assuming that the NCQFT is quantized using the path integral
formulation, as is normally done, it will generate Green functions, covariant in
appearance (if we were  to look upon  $\theta_{\mu\nu} $ as a tensor) and thus
are not expected to coincide with those of (\ref{unitary7}).
So, if we obtain a matrix element of
\begin{equation}
S_{2}(x,y)= i^2T^*[O^*(x)O^*(y)] \label{unitary8}\end{equation} and find that it
differs from that of  $S_{2}(x,y)$ of equation (\ref{unitary7}) [dictated by
causality], we can conclude that causality is violated.
 In other words,
 \begin{eqnarray}
 \Delta & \equiv& T^* [O^* (x)O^* (y)] - T[O^* (x)O^* (y)]
\label{unitary9}\end{eqnarray} can be used to test causality in a quantum field
theory.\\

 We shall now elaborate on $\Delta$ of equation (\ref{unitary9}) and show
that if we had only local interactions or higher order derivative interaction terms
of finite order, $\Delta$ is zero for $x\neq y$. We shall see that for a truly
nonlocal field theory like NCQFT or other nonlocal QFT, $\Delta$ may be
nonzero. Let us consider two operators depending on $\phi(x)$ and its
derivatives:
\begin{eqnarray}
 O_1(x)& \equiv  & D_1\left[ \varphi (x_1 )....\varphi (x_n )\right]\left|_{x_1  = ..... = x_n  = x} \right. \\
 O_2(y)  &\equiv & D_2\left[ \varphi (y_1 )....\varphi (y_n )\right]\left| {_{y_1  = ..... = y_n  = y} } \right.
 \end{eqnarray}
where, $D_1$ and $D_2$ are as yet general  operators that implement
differentiation. Then,
\begin{eqnarray*}
& & T^*[O_1 (x)O_2 (y)] \\
&=& D_1 D_2 [\theta (X^0  - Y^0 )\varphi  (x_1 )....\varphi  (x_n )\varphi  (y_1 )....\varphi  (y_n ) \\
& & + \theta (Y^0  - X^0 )\varphi  (y_1 )....\varphi  (y_n )\varphi  (x_1 )....\varphi  (x_n )]\left| {_{x_1  = ... = x_n  = x,y_1  = ... = y_n  = y} } \right. \\
&=& O_1 (x)O_2 (y)\\
& & + D_1 D_2 \{ \theta (Y^0  - X^0 )[\varphi  (y_1 )....\varphi  (y_n ),\varphi  (x_1 )....\varphi  (x_n )]\} \left| {_{x_1  = ... = x_n  = x,y_1  = ... = y_n  = y} } \right. \\
&=& T[O_1 (x)O_2 (y)] + \theta (y^0  - x^0 )[O_1 (x),O_2 (y)] \\
& & - D_1 D_2 \{ \theta (Y^0  - X^0 )[\varphi  (x_1 )....\varphi  (x_n ),\varphi  (y_1 )....\varphi  (y_n )]\} \left| {_{x_1  = ... = x_n  = x,y_1  = ... = y_n  = y} } \right. \\
\end{eqnarray*}
With \cite{ave},
\[
X^0  = \frac{{\sum\limits_{i = 1}^n {x^0_i } }}{n}~~~  and~~~ Y^0
= \frac{{\sum\limits_{i = 1}^n {y^0_i } }}{n}
\]
Thus,
\[
\begin{array}{l}
 T^*[O_1 (x)O_2 (y)] - T[O_1 (x)O_2 (y)] \\
 = \theta (y^0  - x^0 )[O_1 (x),O_2 (y)] \\
  - D_1 D_2 \{ \theta (Y^0  - X^0 )[\varphi  (x_1 )....\varphi  (x_n ),\varphi  (y_1 )....\varphi  (y_n )]\} \left| {_{x_1  = .... = x_n  = x,y_1  = .... = y_n  = y} } \right. \\
 \end{array}
\]
Suppose $O_{1}$, $O_{2}$ contain finite order time derivatives. Then the above
difference receives contributions only when one or more time derivatives in $D_1$ or $D_2$ act
upon the theta function. This leads to a difference that contains $\delta(x^0-y^0)$
or finite order derivative of $\delta(x^0-y^0)$. The commutators on the other hand
lead to terms $\propto \delta^{3}(\textbf{x-y})$ or its finite order derivatives. The terms
therefore vanish whenever $x\neq y$. On the other hand, for an NCQFT, the
operators $D_1$ and $ D_2$ contain derivatives of an arbitrary order. For example, for
$O_1=O_2= \phi\star \phi\star \ldots \star\phi$,
\[
\begin{array}{l}
 D_1  = e^{\frac{i}{2}\theta ^{\mu \nu } (\partial _\mu ^{x_1 } \partial _\nu ^{x_2 }  + ..... + \partial _\mu ^{x_{n - 1} } \partial _\nu ^{x_n } )}  \\
 D_2  = e^{\frac{i}{2}\theta ^{\mu \nu } (\partial _\mu ^{y_1 } \partial _\nu ^{y_2 }  + ..... + \partial _\mu ^{y_{n - 1} } \partial _\nu ^{y_n } )}  \\
 \end{array}
\]
Such a series acting on $\delta^3(\textbf{x}-\textbf{y})$ smears it over a nonvanishing region in the $x_1-x_2$ plane. This is illustrated in the following section.

\section{Calculations For Causality Violation In NCQFT}

We shall exhibit calculation of  $\Delta$, the difference between $T^*-$product
and T-product, corresponding to the two different cases in the context of two different field theories.
\begin{itemize}
  \item   $ O_1 ^ *  (x) = \frac{{\dot \varphi (x) * \varphi (x) + \varphi (x) * \dot \varphi(x)}}{2}$ in a scalar theory;
  \item $O_2^* (x) = \bar \psi (x)* \psi (x) * \varphi (x)$ in the Yukawa theory
\end{itemize}
While $O^*_1$ cannot be an interaction Lagrangian, being quadratic; this simple example will illustrate the more general case. The essential facet of both the operators is that they contain both a "coordinate" and a "momentum". We shall be considering space-space noncommutativity, i.e. $\theta^{0i}=0$
throughout the calculations.\\
\\
\textbf{Example 1}\\
 Consider the case of the former operator $O_1 ^ *  (x)$. Now,
\begin{eqnarray}
 & &\left\langle {0\left| \Delta  \right|pp'} \right\rangle \nonumber  \\
 & =& \left\langle {0\left| {\left\{ {T^* [:O_1 ^ *  (x)::O_1 ^ *  (y):] - T[:O_1 ^ *  (x)::O_1 ^ *  (y):]} \right\}} \right|pp'} \right\rangle\nonumber  \\
 & =& \left\langle 0\right| \frac{1}{4}\left\{ \partial _0^x \delta (x^0  -y^0 )[:\varphi (y) * \varphi (y ):,:\varphi (x) * \varphi (x ):]\right. \nonumber  \\
 &+&\delta (x^0  - y^0 )[:\varphi (y) * \varphi (y ):,\partial_0^x(: \varphi (x) * \varphi (x ):)]  \nonumber \\
 & -& \left.\delta (x^0  - y^0 )[\partial_0^y(: \varphi (y) * \varphi (y ):),:\varphi (x) * \varphi (x ):] \right\}\left|pp'\right\rangle \label{caus}
 \end{eqnarray}

The right hand side of above equation (\ref{caus}) has three terms. The
commutator in the first term can be Taylor-expanded around $x_0=y_0$. The
leading term (with $x_0=y_0$) is zero for $\theta^{0i}=0$, as shown by
Chaichian et al~\cite{CH02}; and in the second term, we use $(x_0-y_0)\partial _0^x \delta (x^0  -
y^0 )=-\delta (x^0  - y^0 )$. It then cancels second term. Possible nonzero
contribution comes from the third term. Consider the case for which $
\theta^{12}=-\theta^{21}\equiv\theta$, $\theta^{\mu\nu}=0$ otherwise. We find
\begin{eqnarray}
 &&\left\langle {0\left| \Delta  \right|pp'} \right\rangle\nonumber  \\
  &=& \delta (x^0  - y^0 )
(e^{ - ip.x - ip'.y}  + e^{ - ip'.x - ip.y} )\nonumber\\
&\times& \frac{{i}}{{(2\pi
)^{2d - 1} }}\int {d^{d - 1} ke^{i\vec k.(\vec x - \vec y)}
  \cos (\frac{1}{2}\theta ^{ij} k_i p_j )} \cos (\frac{1}{2}\theta ^{ij} k_i p'_j ).\nonumber \\
  &=& (e^{ - ip.x - ip'.y}  + e^{ - ip'.x - ip.y} )\frac{{i}}{{(2\pi )^7 }}\delta (x^0  - y^0 )\nonumber\\
  &\times &\sum\limits_{s =  \pm 1,t =  \pm 1}
  {\delta (x^1  - y^1  - s\theta (p_2  + tp'_2 ))}\nonumber\\
 &&\times\delta (x^2  - y^2  - s\theta (p_1  + tp'_1 ))\delta (x^3  - y^3
 ).\label{css}
 \end{eqnarray}
It may appear that the causality violation term $\Delta$ is non-vanishing only for a specific combinations of coordinate differences and momenta. However, if we use wave-packets for the external lines, there will be a region in the $x_1-x_2-$ plane for which $\Delta$ will be non-zero. \\

\textbf{Example 2}\\ Let us turn to the case of noncommutative Yukawa theory.
\begin{eqnarray*}
S & = & S_{0}+S_{I}\\
S_{0} & = & \int d^{4}x\,\,\left[\bar{\psi}\left[i\partial-m\right]\psi+\frac{1}{2}\left[\partial_{\mu}\phi\partial^{\mu}\phi-m^{2}\phi^{2}\right]\right]\\
S_{I} & = & \lambda\int d^{4}x\,\,\bar{\psi}*\psi*\phi\equiv\lambda\int d^{4}x\,\, L_{I}^{*}\left(x\right)\end{eqnarray*}
 We note that unlike the $\phi^{4}$ theory, the interaction Lagrangian
does contain coordinate $\psi$ and momentum $\bar{\psi}$ at the same time\footnote{However, we note that the integral for $S_{I}$ has no operator ordering
problems.}.  On account of this, the commutator, \[
\left.\left[L_{I}\left(x\right),L_{I}(y)\right]\right|_{x_{0}=y_{0}}\] has terms containing a Dirac
delta-function $\delta^{3}\left(\textbf{\textbf{x-y}}\right)$ and is zero when $x\neq
y$. In a NCQFT, this delta function will get smeared and can be nonzero when
$x_{0}=y_{0},x_{3}=y_{3},x_{\perp}\neq y_{\perp}$.\\
To emphasize the point, we note:\begin{equation} L_I^*(x_0,\textbf{x})L_I^*(x_0,\textbf{y})\neq L_I^*(x_0,\textbf{y})L_I^*(x_0,\textbf{x})\;\;\;\textbf{x}\neq \textbf{y}\end{equation}
We now compute the causality violation amplitude of (\ref{unitary9}) by taking the limit $x_0\rightarrow y_0^+$. We have\footnote{In the present context, the path-integral method does not produce a $T^*-$product in the conventional sense, but is nonetheless symmetric in $x$ and $y$. Hence, we have changed the notation from $T^*\rightarrow\tilde{T}$.},
\begin{eqnarray}
 \Delta _1 & \equiv& \tilde{T} [\bar \psi (x) * \psi (x) * \varphi (x)\bar \psi (y) * \psi (y) * \varphi (y)] \nonumber\\
 &-& T[\bar \psi (x) * \psi (x) * \varphi (x)\bar \psi (y) * \psi (y) * \varphi (y)]\nonumber \\
 &=& \frac{1}{2}\hat D_1 \hat D_2\left\{ [\bar \psi _\alpha  (y)\psi _\alpha  (y_1 ),\bar \psi _\beta  (x)\psi _\beta  (x_1 )]\varphi (x_2 )\varphi (y_2 )\right.\nonumber  \\& +& \left.\mbox{a term involving}\,\left[\phi(x_2),\phi(y_2) \right]\right\}\nonumber
 \end{eqnarray}
With,
\begin{eqnarray}
 \hat{D_1} & = &e^{\frac{i}{2}\theta ^{\mu \nu } [\partial _\mu ^x \partial _\nu ^{x_1 }+
 \partial _\mu ^{x_1} \partial _\nu ^{x_2 }+\partial _\mu ^x \partial _\nu ^{x_2 } ]}\nonumber  \\
 \hat{D_2}  &= &e^{\frac{i}{2}\theta ^{\mu \nu } [\partial _\mu ^y \partial _\nu ^{y_1 }+
 \partial _\mu ^{y_1} \partial _\nu ^{y_2 }+\partial _\mu ^y \partial _\nu ^{y_2 } ]}
\end{eqnarray}
The term involving $\left[\phi(x_2),\phi(y_2) \right]$, (which is a c-number) does not contribute to the following matrix element which we are about to calculate. We now calculate the matrix element of $\Delta_1$ between a state containing two scalars with momenta $l,l'$ and a fermion of momentum $p$  and a state with only a fermion of momentum $p'$.
\begin{eqnarray*}
 2\Delta _2~  & \equiv &2\left\langle {p',s\left|\Delta_1\right|p,s};l,l' \right\rangle  \\
 & =&  \hat{D_1}\hat{D_2} \left\{  \left\langle p',s\right|[\bar \psi_\alpha (y)\psi _\alpha  (y_1 ),\bar \psi _\beta  (x)\psi _\beta(x_1 )] \left|p,s \right\rangle \left\langle {0\left| {\varphi
(x_2 )\varphi (y_2 )} \right|l,l'} \right\rangle\right\}
 \end{eqnarray*}
 $\begin{array}{l}
  = e^{-ilx - il'y} \\
  \times \left[ \begin{array}{l}
 e^{i(p' \wedge l' - p \wedge l)}  \times e^{ip'y - ipx} \delta \left( {(x - y)_i  - \theta ^{ij} p_j  + \theta ^{ij} p'_j  - \theta ^{ij} l_j  + \theta ^{ij} l'_j } \right) -  \\
 e^{i(p' \wedge l - p \wedge l')}  \times e^{ip'x - ipy} \delta \left( {(x - y)_i  + \theta ^{ij} p_j  - \theta ^{ij} p'_j  - \theta ^{ij} l_j  + \theta ^{ij} l'_j } \right) \\
 \end{array} \right]\\\;\;\;\times\bar u^\alpha  (p')\gamma _{\alpha \beta }^o u^\beta  (p) \\
 {\rm{       }} \\
 {\rm{  }} + e^{-il'x - ily} \\
 \times \left[ \begin{array}{l}
 e^{i(p' \wedge l - p \wedge l')}  \times e^{ip'y - ipx} \delta \left( {(x - y)_i  - \theta ^{ij} p_j  + \theta ^{ij} p'_j  + \theta ^{ij} l_j  - \theta ^{ij} l'_j } \right) -  \\
 e^{i(p' \wedge l' - p \wedge l)}  \times e^{ip'x - ipy} \delta \left( {(x - y)_i  + \theta ^{ij} p_j  - \theta ^{ij} p'_j  + \theta ^{ij} l_j  - \theta ^{ij} l'_j } \right) \\
 \end{array} \right]\\\;\;\;\times\bar u^\alpha  (p')\gamma _{\alpha \beta }^o u^\beta  (p) \\
 \end{array}
$ \\
This  nonzero result implies that noncommutative Yukawa theory (which is a nonlocal theory in the sense of nonlocality via the interaction term), is causality violating in the case of space-space noncommutativity (as well as for the space-time noncommutativity).
In this case too, if we consider the matrix elements between the wave-packets states, rather than plane wave states, we will find a  region of causality violation spread over a finite extent in the $x_1-x_2$ plane.

\section{Measurement and the Causality Condition}
We would like to formulate the condition under which two "local" observables in NCQFT, $O_1^*(x)$ and $O^*_2(y)$ are compatible, i.e. their measurements do not interfere with each other. We shall show that this information is already present in the causality condition (\ref{eq:cr}). We shall first consider the  possibility when both $O^*_1=O^*_2=O^*=-iS_1$. Now, the perturbative $U$ matrix differs from identity, $\mathcal{I}$, by the effect of interactions: i.e. it "measures", if indirectly, the impact of an interaction perturbatively. [This is in the same sense that charge density is measured by perturbing the electrostatic potential, or $\theta_{\mu\nu}$ is measured by perturbing the gravitational field $h_{\mu\nu}$]. Thus, $\frac{\delta U}{\delta g(x)}$ measures the effect of observation of $O$ at $x$. $\frac{\delta^2 U}{\delta g(x)\delta g(y)}$ with $x_0<y_0$ has in it the the information of measurement of $O^*(x)$ followed by $O^*(y)$, in the nature of a change in $U$. The effect of measurement of $O(x)$ alone, (i.e. one interaction at $x$ taking place),  on $U$ is to take $U$ from\footnote{Here, $\delta g(x)$ is concentrated around $x$ and we are suppressing an integration.}:
\begin{equation}
\mathcal{I}\rightarrow U= \left(\mathcal{I}+\frac{\delta U}{\delta g(x)}\delta g(x)\right)
\end{equation}
and the effect of measurement of $O^*(y)$ alone is:
\begin{equation}U = \left(\mathcal{I}+\frac{\delta U}{\delta g(y)}\delta g(y)\right)
\end{equation}
The two measurements are compatible if these two "add up" to the net effect of the two successive measurements. In other words, the second order terms,  $\left( O\left[ \delta g(x)\delta g(y)\right]\right)$, in U agree with the compounded effect of two successive measurements:
\begin{eqnarray}
&  &\mbox{second order term in}\,\left(\mathcal{I}+\frac{\delta U}{\delta g(y)}\delta g(y)\right)\left(\mathcal{I}+\frac{\delta U}{\delta g(x)}\delta g(x)\right)\nonumber  \\
& =&\mbox{second order term in}\,\left(\mathcal{I}+\frac{\delta U}{\delta g(x)}\delta g(x)+\frac{\delta U}{\delta g(y)}\delta g(y) \right.\nonumber  \\
& + &\left.\frac{\delta^2 U}{\delta g(x)\delta g(y)}\delta g(x)\delta g(y)\right)
\end{eqnarray}
This can be seen to be just the causality condition (\ref{eq:cr}), expanded  to $O(g^2)$, (recalling $U^\dagger_1=-U_1$ ).
Thus, two observables $O^*(x)$ and $O^*(y)$ are compatible only if
\begin{equation}
\Delta \equiv T^* [O^* (x)O^* (y)] - T[O^* (x)O^* (y)]=0
\label{unitary9'}\end{equation}
This can easily be generalized to the case when arbitrary "local" observables $O^*_1(x)$ and $O^*_2(y)$ are measured. We can introduce sources for $O^*_1(x)$ and $O^*_2(y)$ in the action:
\begin{equation}
S_J=S+\int d^4x\left[ J_1(x)O^*_1(x)+ J_2(x)O^*_2(x)\right]\end{equation}
We can now repeat the above argument, but apply it to $O\left[ \delta J_1(x)\delta J_2(y)\right]$ terms.
Thus, to summarize, "local" observables $O^*_1(x)$ and $O^*_2(y)$ for a NCQFT are compatible only if,
\begin{equation}
\Delta  \equiv T^* [O^*_1 (x)O^*_2 (y)] - T[O^*_1 (x)O^*_2 (y)]=0
\label{unitary9''}\end{equation}

\end{document}